\documentstyle[rawfonts,nassau,pslatex]{article}  
\frompage{000} \topage{000}                                              %| 
\def\rightmark{Systematics of Heavy Quark Production at RHIC} 
\def\leftmark{R. Vogt}
 
\title{Systematics of Heavy Quark Production at RHIC} 
\authors{ 
{R. Vogt}\\[2.812mm]
{\normalsize
Physics Department, University of California, Davis, CA 95616 \\
and\\
Nuclear Science Division,
Lawrence Berkeley National Laboratory \\ 
University of California, Berkeley, California 94720, USA\\[0.2ex] 
}}
 
\abstract{We discuss a program for systematic studies of heavy quark production
in $pp$, $pA$ and $AA$
interactions.  The $Q \overline Q$ production cross sections themselves cannot
be accurately predicted to better than 50\% at RHIC.  For studies of deviations
in $Q \overline Q$ production such as those by nuclear shadowing and heavy
quark energy loss, the $pp$ cross section thus needs to be measured.
We then show that the ratio of $pA$ to $pp$ dilepton mass distributions can
provide a measurement of the nuclear gluon distribution.  With total rates and
nuclear shadowing under control it is easier to study energy loss and to use $c
\overline c$ as a normalization of $J/\psi$ production. }
\keyword{       relativistic heavy-ion collisions,
                heavy flavors
                }

\PACS{  25.75.-q%      Relativistic heavy-ion collisions
}
 
\begin{document}
 
\maketitle
\setcounter{page}{1}

%%%%%%%%%%%%%%%%%%%%%%%%%%%%%%%%%%%%%%%%%%%%%%%%%%%%%%%%%%%%%%%%%%%%%%%%%%%%%%%
\section{Introduction}

It is important to have an accurate measure of the charm and bottom cross 
sections for several reasons.  Heavy quark
decays are expected to dominate the lepton pair continuum from the $J/\psi(c
\overline c)$ and $\Upsilon(b \overline b)$ up to the mass of the $Z^0$ 
\cite{gmrv,losn,cmsdoc}.
Thus the Drell-Yan yield and any thermal dilepton production will essentially
be hidden by the heavy quark decay contributions \cite{gmrv}.  The shape of the
charm and bottom contributions to this continuum could be significantly altered
by heavy quark energy loss \cite{losn,linv}.  
If the 
loss is large, it may be possible to extract a thermal dilepton yield
if it cannot be determined by other means \cite{gkp}.
Heavy quark production in a quark-gluon plasma has also been predicted
\cite{thermc}.  This additional yield can only be determined if the
$AA$ rate can be accurately measured.  Finally, the total charm rate would be
a useful reference for $J/\psi$ production since enhancement of the $J/\psi$ 
to total charm ratio has been predicted in a number of models
\cite{pbmjs,goren12,goren3,grrapp,thewsrhic,thewslhc}.

%%%%%%%%%%%%%%%%%%%%%%%%%%%%%%%%%%%%%%%%%%%%%%%%%%%%%%%%%%%%%%%%%%%%%%%%%%%%%%%
\section{Baseline Rates in $pp$}

We first discuss some new calculations of the $Q \overline Q$ total cross
sections in $pp$ collisions with the most recent nucleon parton distribution
functions.  At leading order (LO) heavy quarks are produced by $gg$ fusion and
$q \overline q$ annihilation while at next-to-leading order (NLO) $qg$ and
$\overline q g$ scattering is also included.  To any order, the partonic 
cross section may be expressed in terms of dimensionless scaling functions
$f^{(k,l)}_{ij}$ that depend only on the variable $\eta$ \cite{KLMV},
\begin{eqnarray}
\label{scalingfunctions}
\hat \sigma_{ij}(\hat s,m_Q^2,\mu^2) = \frac{\alpha^2_s(\mu)}{m^2}
\sum\limits_{k=0}^{\infty} \,\, \left( 4 \pi \alpha_s(\mu) \right)^k
\sum\limits_{l=0}^k \,\, f^{(k,l)}_{ij}(\eta) \,\,
\ln^l\left(\frac{\mu^2}{m_Q^2}\right) \, , 
\end{eqnarray} 
where $\hat s$ is the partonic center of mass energy squared, 
$m_Q$ is the heavy quark mass,
$\mu$ is the scale and $\eta = \hat s/4 m_Q^2 - 1$.  
The cross section is calculated as an expansion in powers of $\alpha_s$
with $k=0$ corresponding to the Born cross section at order ${\cal
O}(\alpha_s^2)$.  The first correction, $k=1$, corresponds to the NLO cross
section at ${\cal O}(\alpha_s^3)$.  It is only at this order and above that
the dependence on renormalization scale, $\mu_R$, enters the calculation
since when $k=1$
and $l=1$, the logarithm $\ln(\mu^2/m_Q^2)$ 
appears.  The dependence on the factorization scale, $\mu_F$, the argument of
$\alpha_s$, appears already at LO.  We assume that $\mu_R = \mu_F = \mu$.  
The next-to-next-to-leading
order (NNLO) corrections to next-to-next-to-leading logarithm
have been calculated near threshold \cite{KLMV} but
the complete calculation only exists to NLO.

The total hadronic cross section is obtained by convoluting the total partonic
cross section with the parton distribution functions (PDFs)
of the initial hadrons,
\begin{eqnarray}
\label{totalhadroncrs}
\sigma_{pp}(s,m_Q^2) = \sum_{i,j = q,{\overline q},g} 
\int_{\frac{4m_Q^2}{s}}^{1} \frac{d\tau}{\tau}\, \delta(x_1 x_2 - \tau) \,
F_i^p(x_1,\mu^2) F_j^p(x_2,\mu^2) \, 
\hat \sigma_{ij}(\tau ,m_Q^2,\mu^2)\, , 
\end{eqnarray}
where the sum $i$ is over all massless partons and
$x_1$ and $x_2$ are fractional momenta.
The PDFs, denoted by $F_i^p$, are evaluated at
scale $\mu$.  All our calculations are fully NLO, applying NLO parton
distribution functions and the two-loop $\alpha_s$ to both the ${\cal
O}(\alpha_s^2)$ and ${\cal O}(\alpha_s^3)$ contributions, as is typically done
\cite{KLMV,MNR}.

\begin{figure}[htb]
%\vspace{40mm}
%\special{psfile=ppcc_mrst.ps %cyl-zB-BW.ps
% vscale=70 hscale=70 voffset=-300 hoffset=-40}
\setlength{\epsfxsize=0.9\textwidth}
\setlength{\epsfysize=0.3\textheight}
\centerline{\epsffile{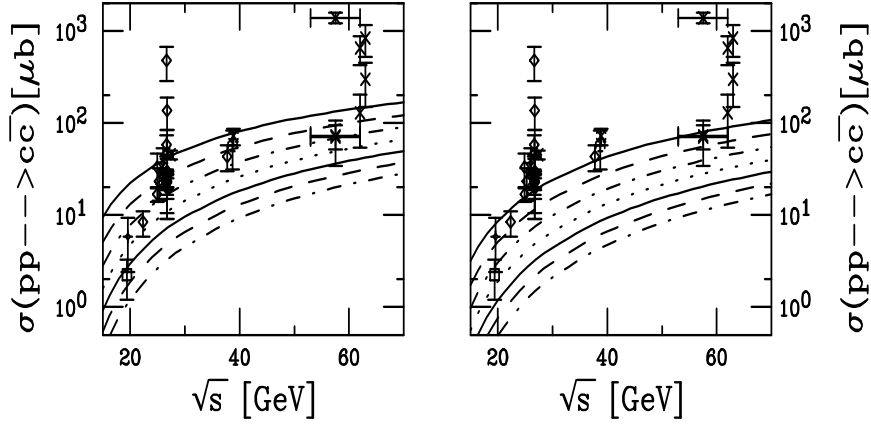}}
\caption[]{Total $c \overline c$ cross sections in $pp$ interactions up to ISR
energies as a function of the charm quark mass.  See \cite{HPC} for references
to the data.  All calculations are fully 
NLO using the MRST HO (central gluon) parton densities.  The left-hand plot 
shows the results with $\mu = m_c$ while in the right-hand plot $\mu = 2m_c$.
From top
to bottom the curves are $m_c = 1.2$, 1.3, 1.4, 1.5, 1.6, 1.7, and 1.8 GeV.}
\label{ppccvmc}
\end{figure}

To obtain the $pp$ cross sections at RHIC and LHC, we first compare the NLO
cross sections to the available $c \overline c$ and $b \overline b$ production
data by varying the mass, $m_Q$, and scale, $\mu$, to obtain the `best'
agreement with the data for several combinations of $m_Q$, $\mu$, and PDF.
We use the recent MRST HO central gluon \cite{mrst}, CTEQ 5M \cite{cteq5}, and
GRV 98 HO \cite{grv98} distributions.  The results for the $c \overline c$
cross section in $pp$ interactions is shown in Fig.~\ref{ppccvmc}.  
%............................................................................
%............................................................................
On the
left-hand side, $\mu = m_c$ for $1.2 \leq m_c \leq 1.8$ GeV, while on the
right-hand side, $\mu = 2m_c$ for the same masses, all calculated with MRST HO.
The scale is not decreased below $m_c$ because the minimum scale in the PDF
is larger than $m_c/2$.  The cross sections with $\mu = m_c$ are all larger
than those with $\mu = 2m_c$ for the same $m_c$ because $\alpha_s(m_c) >
\alpha_s(2m_c)$ by virtue of the running of $\alpha_s$.  Evolution of the PDFs
with $\mu$ tends to go in the opposite direction.  At higher scales the two
effects tend to compensate and reduce the scale dependence but the charm quark
mass is not large enough for this to occur.  

\begin{figure}[tbh]
%\vspace{70mm}
%\special{psfile=cc.ps %cyl-zB-BW.ps
% vscale=70 hscale=70 voffset=-200 hoffset=-40}
% vscale=37 hscale=37 voffset=-95 hoffset=70}
\setlength{\epsfxsize=0.9\textwidth}
\setlength{\epsfysize=0.3\textheight}
\centerline{\epsffile{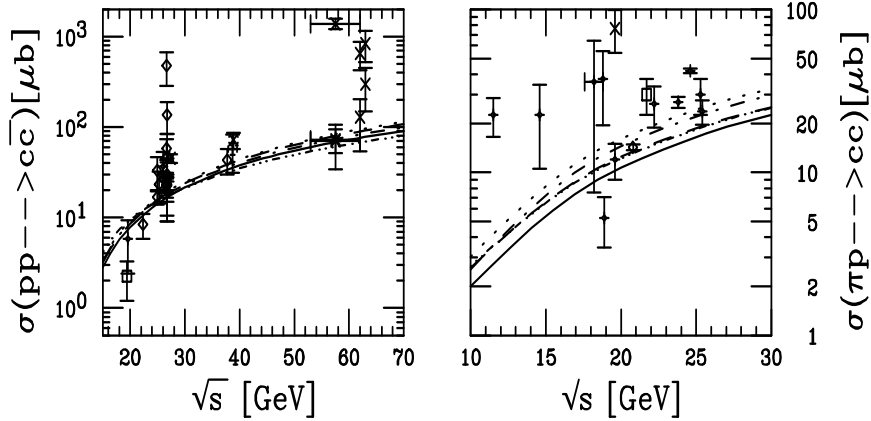}}
\caption[]{Total $c \overline c$ cross sections in $pp$ (left) and $\pi^- p$
interactions compared to data.   See \cite{HPC} for references
(right) to the data.  All calculations are fully 
NLO.  The curves are: MRST HO (central gluon) with $\mu = m = 1.4$ GeV (solid)
and $\mu = 2m = 2.4$ GeV (dashed); CTEQ 5M with $\mu = m = 1.4$ GeV 
(dot-dashed) and $\mu = 2m = 2.4$ GeV (dotted); and GRV 98 HO with $\mu = m =
1.3$ GeV (dot-dot-dot-dashed).}
\label{pppipcc}
\end{figure}

The best agreement with $\mu =
m_c$ is for $m_c = 1.4$ GeV and $m_c = 1.2$ GeV is the best choice for $\mu =
2m_c$ for the MRST HO and CTEQ 5M distributions.  The best agreement with GRV
98 HO is $\mu = m_c = 1.3$ GeV while the results with $\mu = 2m_c$ lie below
the data for all $m_c$.  All five results agree very well with each other for
$pp \rightarrow c \overline c$, as shown on the left side of
Fig.~\ref{pppipcc}.  There is more of a spread in the $\pi^- p \rightarrow c
\overline c$ results, shown on the right side of Fig.~\ref{pppipcc}.  
%............................................................................
%............................................................................
This is
because the $\pi^-$ PDFs are not very well known.  The last evaluations, SMRS
\cite{SMRS}, Owens-$\pi$ \cite{Owenspi}, and GRV-$\pi$ \cite{GRVpi} were 10-15
years ago and do not reflect any of the latest information on the low $x$
behavior of the proton PDFs, {\it e.g.}\ the distributions are all flat as $x
\rightarrow 0$ with no low $x$ rise.  These pion evaluations also depend on the
behavior of the proton PDFs used in the original fit, including the value of
$\Lambda_{\rm QCD}$.  Thus the pion and proton PDFs are generally incompatible.
Note that the $\pi^- p$ cross sections are a bit lower than the data compared
to the $pp$ cross sections, suggesting that lighter quark masses would tend to
be favored for this data.  The reason is because the low $x$ rise in the proton
PDFs depletes the gluon density for $x > 0.02$ relative to a constant at $x
\rightarrow 0$ for $\mu = \mu_0$ the initial scale of the PDF.  The $\pi^- p$
data are in a relatively large $x$ region, $0.1 \leq x = 2\mu/\sqrt{s} \leq
0.3$, where this difference is important.

We have tried to play the same game with the $b \overline b$ total cross
sections but these have mostly been measured in $\pi^- p$ interactions.  The
typical $x$ values of $b \overline b$ production are even larger than those for
$c \overline c$ but it is not clear that $\pi^- p \rightarrow b \overline b$
also favors lower masses.  At the fixed target energies of $b \overline b$
production, $q \overline q$ annihilation dominates while $gg$ fusion is still
most important for $c \overline c$ production \cite{smithrv}.  The
valence-valence $\overline u_\pi u_p$ contribution is most important since
valence distributions dominate at large $x$.  For all three PDFs used, we find
$m_b = \mu = 4.75$ GeV, $m_b = \mu/2 = 4.5$ GeV, and $m_b = 2\mu = 5$ GeV most
compatible with the sparse data.  
Attempts to measure the $b \overline b$ total cross section in fixed-target
$pp$ interactions have been less successful.  Hopefully the HERA-B
experiment at DESY \cite{herab} will soon provide a new measurement.

\begin{figure}[tbh]
%\vspace{40mm}
%\special{psfile=ppcc_lhc.ps %cyl-zB-BW.ps
% vscale=70 hscale=70 voffset=-300 hoffset=-40}
% vscale=37 hscale=37 voffset=-95 hoffset=70}
\setlength{\epsfxsize=0.7\textwidth}
\setlength{\epsfysize=0.3\textheight}
\centerline{\epsffile{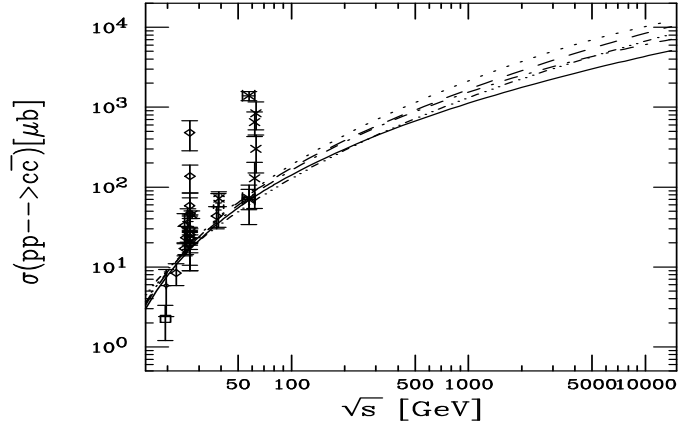}}
\caption[]{Total $c \overline c$ cross sections in $pp$
interactions up to 14 TeV.   See \cite{HPC} for references
to the data.  All calculations are fully 
NLO.  The curves are the same as in Fig.~\ref{pppipcc}.}
\label{ppcclhc}
\end{figure}

Our calculations can then be extrapolated to RHIC and LHC energies.  The result
for $c \overline c$ is shown in Fig.~\ref{ppcclhc}.  
%............................................................................
%............................................................................
Even though the cross
sections agree within 30\% at 40 GeV, by the Pb+Pb energy of the LHC they
differ by a factor of 2.3.  The spread in the $b \overline b$ cross sections is
considerably smaller, $\sim 20-30$\% at the ion collider energies.  Our results
for $pp$ interactions at 40 GeV, 200 GeV, and 5.5 TeV are given in
Table~\ref{qqbtab}.  
\begin{table}[htb]
\begin{center}
\caption{Charm and bottom total cross sections per nucleon for the extrapolated
calculations shown previously.  The heavy quark mass and
factorization/renormalization scales are given, along with the cross
sections at 40 GeV (HERA-B), 200 GeV (Au+Au at RHIC), and 5.5 TeV (Pb+Pb at
LHC).}
\label{qqbtab}
\begin{tabular}{cccccc}
 & & & 40 GeV & 200 GeV & 5.5 TeV \\ \hline
\multicolumn{6}{c}{$c \overline c$} \\
PDF & $m_c$ (GeV) & $\mu/m_c$ & $\sigma$ ($\mu$b) & 
$\sigma$ ($\mu$b) & $\sigma$ (mb) \\ \hline
MRST HO & 1.4 & 1 & 37.8 & 298 & 3.18 \\
MRST HO & 1.2 & 2 & 43.0 & 382 & 5.83 \\
CTEQ 5M & 1.4 & 1 & 40.3 & 366 & 4.52 \\
CTEQ 5M & 1.2 & 2 & 44.5 & 445 & 7.39 \\
GRV 98 HO & 1.3 & 1 & 34.9 & 289 & 4.59 \\ \hline
\multicolumn{6}{c}{$b \overline b$} \\
PDF & $m_b$ (GeV) & $\mu/m_b$ & $\sigma$ (nb) & 
$\sigma$ ($\mu$b) & $\sigma$ ($\mu$b) \\ \hline
MRST HO & 4.75 & 1 & 9.82 & 1.90 & 185.2 \\
MRST HO & 4.5  & 2 & 8.73 & 1.72 & 193.2 \\
MRST HO & 5.0  & 0.5 & 10.96 & 2.16 & 184.8 \\
GRV 98 HO & 4.75 & 1 & 13.40 & 1.65 & 177.6 \\
GRV 98 HO & 4.5  & 2 & 12.10 & 1.64 & 199.0 \\
GRV 98 HO & 5.0  & 0.5 & 14.80 & 1.73 & 166.0 \\ \hline
\end{tabular}
\end{center}
\end{table}
The $AA$ rates per event at $b =0$ with the same 
energies can be obtained by multiplying these cross sections by $T_{AA}(b=0)$,
29.3/mb for Au+Au and 30.4/mb for Pb+Pb.  We find 8-13 $c \overline c$ pairs
and $\sim 0.05$ $b \overline b$ pairs at RHIC with 97-225 $c \overline c$ pairs
and $\sim 5$ $b \overline b$ pairs at LHC without nuclear shadowing.  The
shadowing effect is rather small for $c \overline c$ at RHIC and actually
enhances the $b \overline b$ rate.  The only important modification due to
shadowing in the total cross section is on the $c \overline c$ rate at the LHC
which is reduced to 67-150 pairs.  Energy loss does not affect the
total rate \cite{linv}.  As noted by Thews, this $c \overline c$ rate is 
large enough at
RHIC for independent $c$ and $\overline c$ quarks to dynamically recombine to
form $J/\psi$'s \cite{thewsrhic}.  
The baseline rates of $Q \overline Q$ production are thus
important for studying these effects.

%%%%%%%%%%%%%%%%%%%%%%%%%%%%%%%%%%%%%%%%%%%%%%%%%%%%%%%%%%%%%%%%%%%%%%%%%%%%%%%
\section{Nuclear Gluon Distribution in $pA$}

We now turn to a calculation of the nuclear gluon distribution in $pA$
interactions \cite{ekv}.  We show that the dilepton continuum can be used to
study nuclear shadowing and reproduces the input shadowing function well, in
this case, the EKS98 parameterization \cite{EKS}.  To simplify
notation, we refer to generic heavy quarks, $Q$, and heavy-flavored mesons,
$H$.  The lepton pair production cross section is
\begin{eqnarray}
  \frac{d \sigma^{pA \rightarrow l \overline l +X}}
  {dM_{l \overline l} dy_{l \overline l}} & = &
   \int d^3\vec p_{l} d^3\vec p_{\overline l} 
                 \int d^3\vec p_{H} d^3\vec p_{\overline H}
    \,\delta(M_{l \overline l}-M(p_l,p_{\overline l})) 
    \,\delta(y_{l \overline l}-y(p_l,p_{\overline l}))  \nonumber \\
  & & \times
    \frac{d\Gamma^{H \rightarrow l+X}(\vec p_H)}{d^3 \vec p_l}
    \,\, 
    \frac{d\Gamma^{\overline H \rightarrow \overline l+X}(\vec p_{\overline 
     H})}
         {d^3 \vec p_{\overline l}}
    \,\, 
    \frac{d\sigma^{{\rm p}A \rightarrow H \overline H+X}}
         {d^3 \vec p_{H} d^3 \vec p_{\overline H }} \qquad \qquad
    \nonumber \\
  & & \times 
    \theta(y_{\rm min}<y_{l},y_{\overline l}<y_{\rm max})
    \theta(\phi_{\rm min}<\phi_{l},\phi_{\overline l}<\phi_{\rm max})  
   \qquad \label{dspAll}
\end{eqnarray}
where $M  (p_l,p_{\overline l})$ and $ y(p_l,p_{\overline l})$ are the
invariant mass and rapidity of the $l \overline l$ pair.
The decay rate, $d\Gamma^{H \rightarrow l+X}(\vec p_H)/d^3 \vec p_l$, is 
the probability that meson $H$ with momentum $\vec p_H$ decays
to a lepton $l$ with momentum $\vec p_l$.  The $\theta$ functions define
single lepton rapidity and azimuthal angle cuts used to simulate
detector acceptances.

Using a fragmentation function $D^{H}_{Q}$ to describe quark
fragmentation to mesons, the $H \overline H$ production cross section 
can be written as
\begin{eqnarray}
  \frac{d \sigma^{pA \rightarrow H \overline H+X}}
       {d^{3} \vec p_H d^3 \vec p_{\overline H}}  &= & 
    \int \frac{d^3\vec p_{Q}}{E_{Q}} 
         \frac{d^3\vec p_{\overline Q}}{E_{\overline Q}}
   \, E_{Q}E_{\overline Q}\frac{d\sigma^{pA \rightarrow Q 
    \overline Q+X}}
        {d^{3} \vec p_Q d^3 \vec p_{\overline Q}}  
    \int dz_1 dz_2 D^{H}_{Q}(z_1)
     D^{\overline H}_{\overline Q}(z_2) \nonumber \\
   & \times & \, \delta^{(3)}(\vec p_H - z_1 \vec p_Q)
             \, \delta^{(3)}(\vec p_{\overline H} - z_2 \vec p_{\overline Q})
\, \, . \label{dspADD}
\end{eqnarray}
Our calculations were done with two different fragmentation functions.  We
found that our results were independent of $D^H_Q$.
The hadronic heavy quark production cross section {\em per nucleon} in $pA$
collisions can be factorized into the general form
\begin{eqnarray}
\frac{1}{A} E_{Q}E_{\overline Q} \frac{d\sigma^{pA \rightarrow
   Q \overline Q + X}}
    {d^{3} \vec p_Q d^3 \vec p_{\overline Q}} = \sum_{i,j}
    \int dx_1 dx_2 \, f_i^p(x_1,\mu^2) f_j^A(x_2,\mu^2) 
     E_{Q}E_{\overline Q}
    \frac{d\hat \sigma^{ij \rightarrow Q \overline Q}}
                       {d^3 \vec p_{Q} d^3 \vec p_{\overline Q}}
\end{eqnarray}
where $f_i^p = F_i^p/x$ and $f_i^A = F_i^A/x$ with $F_i^A = F_i^p R_i^A$.
The shadowing ratio $R_i^A$ is that of EKS98 \cite{EKS}.  The partonic cross
section is the differential of Eq.~(\ref{scalingfunctions}) at $k=0$.
%Integration of Eqs.~(\ref{dspAll}) and (\ref{dspADD}) over the total
%phase space gives the normalization $\sigma^{{\rm p}A \rightarrow l
%\overline l} = B^2 \sigma^{{\rm p}A \rightarrow H \overline H} = B^2
%\sigma^{{\rm p}A \rightarrow Q \overline Q}$, where
%$B=\Gamma_l/\Gamma$.  
Note that the total lepton pair production cross section is equal to the total
$Q \overline Q$ cross section multiplied by the square of the 
lepton branching ratio.

\begin{figure}[htb]
\setlength{\epsfxsize=0.5\textwidth}
\setlength{\epsfysize=0.5\textheight}
\centerline{\epsffile{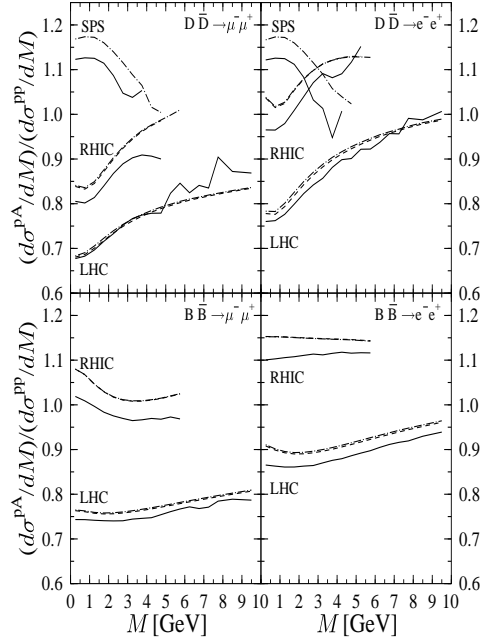}}
%\vspace{40mm}
%\special{psfile=FigdsRg2.eps %cyl-zB-BW.ps
% vscale=37 hscale=37 voffset=5 hoffset=70}
\caption[]{The ratios of lepton pairs from correlated $D \overline D$ and
$B \overline B$ decays in $pA$ to $pp$ collisions at the same energies (solid
curves) compared to the input $R_g^A$ at the
average $x_2$ and $\mu$ (dashed)/$\sqrt{\langle \mu^2 \rangle}$ (dot-dashed)
of each $M$ bin.  From Ref.~\cite{ekv}.
}
\label{lepvsm}
\end{figure}

We compare the ratios of lepton pair cross sections with
the input $R_g^A$ in Fig.~\ref{lepvsm}. 
All the results are
integrated over the rapidity intervals appropriate to the PHENIX and ALICE
dilepton coverages.  The ratio follows $R_g^A$ at all
energies. The higher the energy, the better the agreement: at the LHC
the two agree very well.

%............................................................................
%............................................................................
The ratio always lies below $R_g^A$ for two reasons.
First, $q \overline q$ annihilation is included and quark shadowing
is different than gluon shadowing.  The $q
\overline q$ contribution decreases with energy, leading to better
agreement at the LHC.
Second,
the phase space integration smears the shadowing effect relative to
$R_g^A(\langle x_2 \rangle, \langle \mu \rangle)$.  
Note that the ratio deviates slightly more from $R_g^A(\langle
x_2\rangle,\langle \mu\rangle)$ for $e^+ e^-$ than for $\mu^+ \mu^-$ 
because the curvature of $R_g^A$ with
$x$ is stronger at larger values of $x$ and, due to the differences in rapidity
coverage, the average values of $x_2$ are larger for
$e^+ e^-$.

The average $x_2$ decreases with energy.  We have
$0.14 \leq \langle x_2 \rangle
\leq 0.32$ at the SPS where $R_g^A$ is decreasing.  
At RHIC,  $0.003 \leq \langle x_2 \rangle \leq
0.012$, where $R_g^A$ is increasing quite rapidly. Finally, at the
LHC, $3\times 10^{-5} \leq \langle x_2 \rangle \leq 2\times 10^{-4}$
where $R_g^A$ is almost independent of $x$. The values of
$\langle x_2 \rangle$ are typically larger for electron pairs at
collider energies because the electron coverage is more central than
the muon coverage.

The average $\mu^2$ increases with energy and quark mass. For $c\overline c$ 
we have $7.58 \leq
\langle \mu^2 \rangle \leq 48.5$ GeV$^2$ at the SPS, $9.46 \leq \langle \mu^2
\rangle \leq 141$ GeV$^2$ at RHIC, and $11.4 \leq \langle \mu^2 \rangle
\leq 577$ GeV$^2$ at the LHC.  For $b \overline b$ production, 
$32.0 \leq \langle \mu^2
\rangle \leq 54.3$ GeV$^2$ at RHIC and $37.9 \leq \langle \mu^2 \rangle \leq
156$ GeV$^2$ at LHC.

\section{Heavy Quarks in $AA$}

\subsection{Effects of Energy Loss}

Energy loss would best be determined by reconstruction of $D$ and $B$ meson
decays and comparing with distributions expected from $pp$ and $pA$
extrapolations that do not consider energy loss.  Whether energy loss is
measurable in reconstructed $D$ and $B$ decays or not, the change in the
dilepton continuum should surely be present if the loss is nonzero and will
bias the interpretation of the dilepton continuum.
So far, the amount of the energy
lost by heavy quarks is unknown.  While a number of calculations have been 
made of the collisional loss in a quark-gluon plasma \cite{coll}, only 
recently has radiative loss been applied to heavy quarks \cite{mtdks}.  The
radiative loss can be rather large, $dE/dx \sim -5$ GeV/fm for a 10 GeV
heavy quark, and
increasing with energy, but the collisional loss is smaller, $dE/dx \sim 1-2$
GeV/fm, and nearly independent of energy \cite{mtdks}.  
We note that energy loss
will suppress high $p_T$ and large invariant mass quark pairs as long as
$|dE/dx| \geq \langle p_T 
\rangle/R_A$ \cite{linv}.

It is important to note that energy loss does not reduce the number of $Q
\overline Q$ pairs produced but only changes their momentum.  However, an 
effective reduction in the observed heavy quark yield 
can be expected in a finite acceptance detector because
fewer leptons from the subsequent decays of the heavy quarks will pass 
kinematic cuts such as a minimum lepton $p_T$.  

If the loss or the $p_T$ cut
is large, the Drell-Yan and thermal dileptons could emerge from
under the reduced $D \overline D$ and $B \overline B$ decay contributions 
at large masses.
Even without considering energy loss, Gallmeister {\it et al.}\ suggested that
thermal dileptons could be detected by increasing the minimum lepton $p_T$
because, in the $D$ and $B$ rest frames, the maximum energy of the
individual leptons is limited to 0.9 and 2.2 GeV respectively.  The lepton 
$p_T$ from thermal production has no such limitation \cite{gkp}.

\subsection{Quarkonium normalization}

Heavy quark production in $AA$ collisions is also interesting because of the
prominent effect it could have on quarkonium.
Initial nucleon-nucleon collisions may not be the only source of quarkonium
production.  Regeneration of quarkonium in the
plasma phase \cite{pbmjs,goren12,goren3,grrapp,thewsrhic,thewslhc}
could counter the effects of 
suppression, ultimately leading to enhanced quarkonium production.  In the
plasma phase, there are two basic approaches: statistical and dynamical
coalescence.  Both these approaches depend on being able to measure the
quarkonium rate relative to total $Q \overline Q$ production.
Thus the $Q \overline Q$ rate is preferable as a normalization of quarkonium
production, particularly since both share the same production mechanisms and
approximate $\langle x \rangle$, $\langle \mu \rangle$ values.  However the
final-state effects such as energy loss will make the total rate difficult to 
quantify without substantial detailed studies.
These secondary production models should be testable already at RHIC where
enhancements of factors of 2-3 are expected from coalescence 
\cite{goren3,thewsrhic}.  

Other processes besides heavy quark production
have been suggested as references for quarkonium production.
Using the $Z^0$ as a reference \cite{RVz} as a reference would eliminate the
uncertainty due to final-state effects on the $Z^0 \rightarrow l^+l^-$ decays
but the different production mechanisms and masses leaves it less desirable.
It has also been suggested that the $\psi'/J/\psi$ and $\Upsilon'/\Upsilon$
ratios be studied as a function of $p_T$ \cite{GunV} since deviations from the
$pp$ ratios should reflect quark-gluon plasma characteristics.  The only
drawback to such a mechanism is that strong suppression may result in poor
statistics. 

%%%%%%%%%%%%%%%%%%%%%%%%%%%%%%%%%%%%%%%%%%%%%%%%%%%%%%%%%%%%%%%%%%%%%%%%%%%%%%%

%%%%%%%%%%%%%%%%%%%%%%%%%%%%%%%%%%%%%%%%%%%%%%%%%%%%%%%%%%%%%%%%%%%%%%%%%%%%%%%

\section*{Acknowledgements}
I would like to thank K.J. Eskola and V.J. Kolhinen for collaboration on
part of this work.  
This work was supported by the Director,
Office of Energy Research,
Office of High Energy and Nuclear Physics,
Nuclear Physics Division of the U.S.\ Department of Energy
under Contract No.\ DE-AC03-76SF00098.

%%%%%%%%%%%%%%%%%%%%%%%%%%%%%%%%%%%%%%%%%%%%%%%%%%%%%%%%%%%%%%%%%%%%%%%%%%%%%%%

%%%%%%%%%%%%%%%%%%%%%%%%%%%%%%%%%%%%%%%%%%%%%%%%%%%%%%%%%%%%%%%%%%%%%%%%%%%%%%%
\vfill
%\noindent
%{\small \sl 18$^{\rm th}$ Winter Workshop on Nuclear Dynamics, 21-26 January
%2002, Nassau, The Bahamas.}
\eject
\end{document}